\documentclass[preprint,authoryear,1p]{elsarticle}

\usepackage{amsmath}
\usepackage{amssymb}
\usepackage{bm}
\usepackage{caption}
\usepackage{subcaption}
\usepackage{url}
\usepackage{color}

\newcommand{\unit}[1]{\,\textrm{#1}}

\journal{ }

\journal{IJNME}
\begin{document}

\begin{frontmatter}



\title{Topology optimization of microwave waveguide filters}

\author[mekn]{N. Aage\corref{cor1}}
\ead{naage@mek.dtu.dk}
\author[mekn,cam]{V.Egede Johansen}
\address[mekn]{Department of Mechanical Engineering, Solid Mechanics, Technical University of Denmark, Bld. 404, DK-2800 Kgs. Lyngby.}
\address[cam]{Department of Chemistry, University of Cambridge, Lensfield Rd., Cambridge, CB21EW, UK}

\cortext[cor1]{Corresponding author}

\begin{abstract}
We present a density based topology optimization approach for the design of metallic microwave insert filters. A two-phase optimization procedure is proposed in which we, starting from a uniform design, first optimize to obtain a set of spectral varying resonators followed by a band gap optimization for the desired filter characteristics. This is illustrated through numerical experiments and comparison to a standard band pass filter design. It is seen that the carefully optimized topologies can sharpen the filter characteristics and improve performance. Furthermore, the obtained designs share little resemblance to standard filter layouts and hence the proposed design method offers a new design tool in microwave engineering. 
\end{abstract}


\end{frontmatter}


\section{Introduction}
Microwave filters are crucial to modern day telecommunication, sensing and other high technology applications. Ever since Marconi's (wideband) spark-gap transmitters were replaced by continuous wave transmitters and the radio frequency spectrum started to crowd, microwave filters have been key in improving receiver performance and avoiding signal leaking at unwanted frequencies.

One class of microwave filters are made to fit metallic waveguides. Metallic waveguides  are tube-like structures -- most often of rectangular cross-section -- inside which waves can propagate. They are low-loss and therefore suitable for a palette of applications ranging from extremely sensitive systems (like space communication) or high energy systems where losses would lead to heating or significant energy waste (radars, RF based particle accelerators, microwave ovens).

Waveguide filter design normally consists of designing a prototype filter network and then cascading inductive, capacitive or resonating elements to implement the prototype filter \citep{Pozar2005}. Some elements extend the geometry of the waveguide by coupling the waveguide to external resonators or stubs. Other elements are placed inside the existing waveguide geometry, like so-called post filters, iris coupled filters, dielectric resonators and insert filters. These filter design elements are usually based on structures to which an approximate analytical model exists, such that the behavior can be shaped to fit the prototype filter elements. This enables realization of almost any desired filter, but often at the practical cost of vast space/weight consumption due to cascading of elements.

To enable more compact, high-performance filters, this work proposes a topology optimization approach to filter design. Such approach has the clear advantage of being able to design \emph{one} integrated structure to obtain a given set of filter properties instead of a cascaded series of elements. 
Former work on synthesis of waveguide filters using topology optimization has mainly focused on dielectric materials \citep{Byun2007,Khalil2008a,Khalil2008b,Khalil2009,Choi2012} which can be fabricated using e.g. additive manufacturing techniques \citep{Delhote2011}. Sharper filter responses can in general be obtained using metallic structures, but topology optimization of metallic structures is a much more difficult problem, mainly due to the non-monotonic material interpolation between metal and air \citep{Aage2010}.

The only successful attempt on topology optimization of metallic waveguide filters known by the authors is by \citet{Quedraogo2012}, but here a genetic algorithm was used which strongly limits its application to larger problems. Other attempts on metallic micro strip resonator design also exists \citep{Assadihaghi2006} but no usable designs were obtained. Interest in research on metallic microwave optimization is increasing \citep{Aage2010,Erentok2011,Hassan2014,Nomura2013,Hassan2015} and the methods has been applied several times to antenna design problems. Using experience and knowledge gained through these works, we are at a state where attempting to design usable metallic waveguide filters is a feasible task.


%
%
%

\section{Physical model}
\begin{figure}
\centering
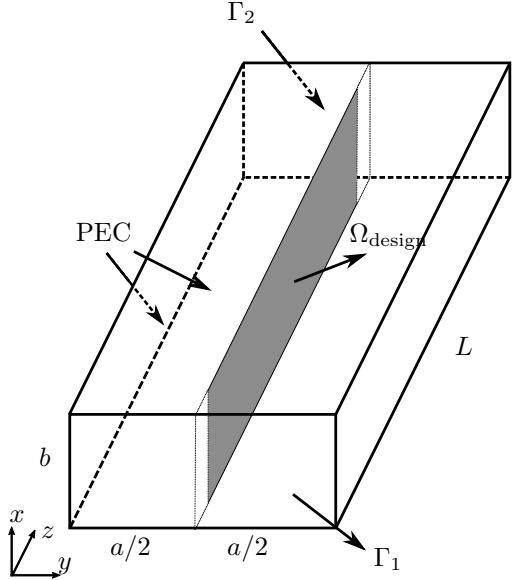
\caption{Basic geometry for the rectangular waveguide problem (WR-90) with $a=22$mm and $b=10$mm. The boundary condition, i.e. PEC and waveguide ports can be seen as well as the design domain for the optimization process, here denoted with $\Omega_{\text{design}}$.}
\label{fig:problemsketch}
\end{figure}
The model problem to be investigated in this is illustrated in Fig. \ref{fig:problemsketch}. The figure shows a section of a rectangular waveguide in which we need to solve Maxwell's equations along with appropriate boundary conditions \citep{balanis2012,jin2002}. 
In this work, it is sufficient to assume linear, isotropic materials with no free charges and solve the problem in the frequency domain. Maxwell's equations are therefore conveniently cast as a single second order Partial Differential Equation (PDE) in the electric field as follows
\begin{equation}
\label{eq:maxeqs_E} \nabla \times \left( \mu_r^{-1} \nabla \times \bm{E} \right) - k_0^2 \left( \epsilon_r - j \frac{\sigma}{\omega \epsilon_0}\right) \bm{E} = 0 ~ \textup{in} ~ \Omega,
\end{equation}
where $\bm{E}=\bm{E}(\bm{x})$ is the unknown complex electric field, $\omega$ is the frequency, $\mu_r$ is the relative permeability, $\epsilon_r$ the relative permittivity, $\sigma$ is the conductivity and $k_0 = \omega \sqrt{\epsilon_0 \mu_0}$
is the free space wave number. The two remaining
constants are the free space permittivity $\epsilon_0\approx 8.8542\cdot 10^{-12} \unit{F/m}$ 
and the free space permeability $\mu_0 = 4 \pi 10^{-7} \unit{H/m}$. The real, physical solution can then be obtained from $\bm{E}_{\text{phys}}(\bm{x},t)=\mathbb{R}(\bm{E}(\bm{x}) e^{j \omega t})$.

The vector wave equation from Eq. \eqref{eq:maxeqs_E} is solved for the following boundary conditions in the case of a metallic waveguide. 
The four sides that encloses the waveguide are modeled as Perfect Electric Conductors (PEC) which means that the tangential component of the
electric field is zero, i.e.
\begin{equation}
\label{eq:PEC}
\bm{n} \times \bm{E}  = \bm{0} ~~ \text{on} ~~ \Gamma_{\text{PEC}},
\end{equation}
where $\bm{n}$ is an outward normal to the respective surfaces. The input and output ports, denoted with $\Gamma_1$ and $\Gamma_2$ in Fig. \ref{fig:problemsketch}, are modeled using a waveguide port boundary condition \cite{jin2002}. This boundary condition utilizes that 
only certain modes can propagate in a given waveguide geometry and as a consequence these modes can be used to excite and absorb the electric field passing through the boundary. The modes can be solved analytically for simple geometries, or numerically for complicated port geometries. For the rectangular waveguide the first transverse electric mode, $\text{TE}_{10}$, is
\begin{equation}
\label{eq:TE10}
\bm{e}_{10}^{\text{TE}}(\bm{x}) =  - \sqrt{\frac{2}{ab}}\frac{\pi^2}{a^2}
 \begin{Bmatrix}
 0 \\
\sin(\pi x / a) \\
 0
 \end{Bmatrix},
\end{equation}
when the port is situated in the $xy$-plane (i.e. the wave is propagating along the $z$-axis). Note that it is common practice to operate waveguides using only one mode. Now the boundary condition for the input and output port can be stated as
\begin{align}
\label{eq:ports_1}
\bm{E} & = \bm{E}_{\text{inc}} + c_{10,1} \bm{e}_{10}^{\text{TE}} e^{j \gamma z} &\text{on} ~~ \Gamma_{1},\\
\label{eq:ports_2}
\bm{E} & =  c_{10,2} \bm{e}_{10}^{\text{TE}} e^{j\gamma z}~~ &\text{on} ~~ \Gamma_{2},
\end{align}
where the propagation constant $\gamma = \sqrt{k_z^2 - k_0^2}$ and $k_z = \pi / a$ is the port cutoff wave number. The constants $c_{10,i}$ can be determined from orthogonality. The latter yields the following
\begin{align}
\label{eq:constants}
c_{10,1} & =  e^{ - j \gamma z}\int_{\Gamma_{1}} \bm{e}_{10}^{\text{TE}} \cdot
\left( \bm{E} - \bm{E}_{\text{inc}} \right)~ \text{d}\Gamma ,
 \\
c_{10,2} & =  e^{ - j \gamma z}\int_{\Gamma_{2}} \bm{e}_{10}^{\text{TE}} 
\cdot \bm{E} ~\text{d}\Gamma .
\end{align}
These expressions are now re-inserted into Eqs. \eqref{eq:ports_1} and \eqref{eq:ports_2} after which the natural (Neumann) operator for the Maxwell problem, i.e. $\bm{n} \times \nabla \times $, is applied. Furthermore, the incident wave is chosen as the first order mode for the waveguide. Combining all of the above we obtain the following port boundary conditions for the waveguide problem
\begin{align}
\label{eq:ports_1_BC}
\bm{n} \times \nabla \times \bm{E} & = j \gamma \bm{e}_{10}^{\text{TE}}  \cdot \int_{\Gamma_{1}} \bm{e}_{10}^{\text{TE}} \cdot  \bm{E} ~\text{d}\Gamma
- 2 j \gamma \bm{e}_{10}^{\text{TE}}  &\text{on} ~~ \Gamma_{1},\\
\label{eq:ports_2_BC}
\bm{n} \times \nabla \times \bm{E} & = j \gamma \bm{e}_{10}^{\text{TE}} \cdot \int_{\Gamma_{2}} \bm{e}_{10}^{\text{TE}} \cdot  \bm{E} ~\text{d}\Gamma ~~ &\text{on} ~~ \Gamma_{2},
\end{align}
which can naturally be incorporated into a variational form for the Maxwell PDE. 
\section{Finite element formulation and implementation}
The PDE problem described in Eqs. \eqref{eq:maxeqs_E}, \eqref{eq:PEC}, \eqref{eq:ports_1_BC} and \eqref{eq:ports_2_BC} leads to the following linear finite element formulation when using the standard Galerkin discretization approach \citep{jin2002}:
\begin{equation}
\label{eq:FEM}
\bm{S}(\omega) \bm{E} = \left( \bm{K} - \bm{M} +  \bm{B}_{\Gamma_1} + \bm{B}_{\Gamma_2} \right) \bm{E} = \bm{f}(\omega),
\end{equation}
where the global system matrices and vectors are assembled from
\begin{align}
\label{eq:element_K}
\bm{K} &= \sum\limits_{e}\int_{\Omega_e} \mu_r^{-1} (\nabla \times \bm{N}_e)^T (\nabla \times \bm{N}_e)\text{d}\Omega, \\
\label{eq:element_M}
\bm{M} &= \sum\limits_{e}\int_{\Omega_e}  k_0^2 \left( \epsilon_r - j \frac{\sigma}{\omega \epsilon_0}\right) \bm{N}_e^T \bm{N}_e \text{d}\Omega, \\
\label{eq:element_B}
\bm{B}_{\Gamma_i} &= j \gamma \sum\limits_{e} \int_{\Gamma_i} \bm{N}_e^T  \bm{e}_{10}^{\text{TE}} \text{d}\Gamma \cdot \sum\limits_{e} \int_{\Gamma_i}   \left(\bm{e}_{10}^{\text{TE}}\right)^T \bm{N}_e \text{d}\Gamma, ~ i = 1,2, \\
\label{eq:element_f}
\bm{f} & = 2 j \gamma \sum\limits_{e} \int_{\Gamma_1} \bm{N}_e^T  \bm{e}_{10}^{\text{TE}} \text{d}\Gamma, 
\end{align}
where $\bm{N}(\bm{x})$ represent vector (Nedelec) element shape functions belonging to the Sobolev space 
$V:=H(\nabla \times \bm{v}, \Omega)=\{ \bm{v} \in [L_2(\Omega)]^3 ~| ~ \nabla \times \bm{v} \in [L_2(\Omega)]^3 \}$. In this work we use tetrahedral elements with zero/first order shape functions. 

\begin{figure}
\centering
\includegraphics[width=0.8\textwidth]{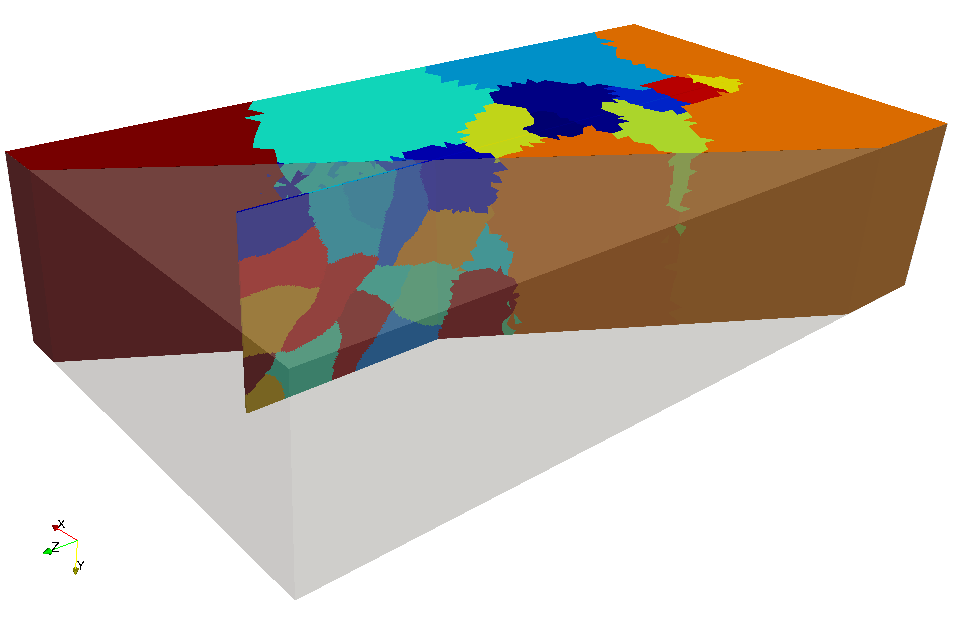}
\caption{Partitioning for the waveguide problem. Note that the port is required to be on a single partition to facilitate easy construction
of the outer product for the boundary integrals in Eq. \eqref{eq:element_B}. This can be seen by the orange partition to the right that fully encloses the 2nd port surface. The thin plate in the center is the design domain.}
\label{fig:parts}
\end{figure}

The finite element problem in Eqs. \eqref{eq:FEM} - \eqref{eq:element_f} is implemented in an in-house MPI based C++ code \citep{Aage2013b}. The meshing is performed in Cubit (Sandia National Laboratories) and the partitioning is done by METIS \citep{Karypis1999}. It is important to note that the outer product in Eq. \eqref{eq:element_B} can result in a high degree of communication in the parallel implementation in case the port boundary is distributed which is due to the fact that the $\bm{B}$ matrices are dense. Therefore we require that each port is confined to a single partition as is illustrated in Fig. \ref{fig:parts}. The figure also indirectly shows the design domain as a thin plate which is partitioned heavily due to the fine mesh desired for a fine design representation. The Maxwell problem is finally solved using the direct solve MUMPS \citep{Amestoy2000}.
\section{Design interpolation}
The topology optimization methodology presented next falls into the category of so-called density based approaches \citep{bendsoe2003}. That is,
for each element in the design domain we assign a continuous variable, i.e. $\rho_e \in [0;1]$ which we use to interpolate between two candidate materials. In the case of metallic waveguide filters, the relevant physical quantity is the conductivity. It is desired that the final design does not contain intermediate valued design variable such that the design is physically realizable. To obtain an interpolation scheme with such properties we adopt the scheme presented in \citet{Aage2010}. For completeness the interpolation scheme is stated below.

The design interpolation consists of two parts. The first interpolates directly in the material conductivity in the original Maxwell problem, i.e. Eq. \eqref{eq:maxeqs_E} or \eqref{eq:FEM} and can be written as
\begin{equation}
\label{eq:sigma}
\sigma(\rho) = \sigma_0 10^{\left[\log_{10}\left(\frac{\sigma_d}{\sigma_0}\right) + \rho \left\{ 
\log_{10}\left(\frac{\sigma_m}{\sigma_0}\right) - \log_{10}\left(\frac{\sigma_d}{\sigma_0}\right)
 \right\} \right]},
\end{equation}
where subscripts $m$ and $d$ refers to the metal and the background dielectric respectively, and where $\sigma_0$ is a scaling included to make the interpolation physically sound. In the work presented here we use $\sigma_0= \sigma_m=10^6 \unit{S/m}$ and $\sigma_d=10^{-4} \unit{S/m}$. The given values are chosen for the following reasons. Due to numerical precision it can be shown that using a conductivity larger than $10^6 \unit{S/m}$ for the metal does not change the response of the system  significantly.
Using a larger value (e.g. copper has $\sigma_m=5.998 \times 10^6\unit{S/m}$) will  lead to intermediate design variables in the final design since the system performance does not benefit from an increase in the conductivity above the proposed upper bound. 
The same argument holds for the background material, i.e. choosing $\sigma_d$ too low leads to numerical instabilities, while a too large value results in an over damped response and hence a non-monotonic behaving design interpolation. 

The second part of the interpolation scheme takes the skin depth issue into account. That is, the distance an electromagnetic wave propagates into a good conductor before being reduced by a factor of $e^{-1}$. 
Since the skin depth of a good conductor is several orders of magnitude smaller than the actual device dimensions, the finite element model cannot capture the rapid 
decay unless additional measures are taken. If this is not treated the skin depth effect will again result in a non-monotonically behaving design interpolation on meshes that resolve the geometry but not the skin depth. To circumvent this limitation we apply an impedance boundary condition on all faces of a design element. This leads to the following condition
\begin{equation}
\label{eq:skindepth}
\bm{n} \times \nabla \times \bm{E} + \rho_e^{13} j k_0 \sqrt{\frac{\epsilon_m - j \frac{\sigma_m}{\omega \epsilon_0}}{\mu_m}} 
\bm{n} \times \bm{n} \times \bm{E} = 0 ~ \text{on} ~ \Gamma_e,
\end{equation}
where $\Gamma_e$ refer to \textit{all} faces of the $e$'th design element. From the two Eqs. \eqref{eq:sigma} and \eqref{eq:skindepth} it is clear that $\rho_e=1$ yields a good conductor and that $\rho_e=0$ results in a background dielectric with very little artificial damping.
For further details on the interpolation scheme and its parameters the reader is referred to \cite{Aage2010}.
\section{Objective function}
\begin{figure}[htb]
  \begin{subfigure}[b]{0.47\textwidth}
	 \includegraphics[width=\textwidth]{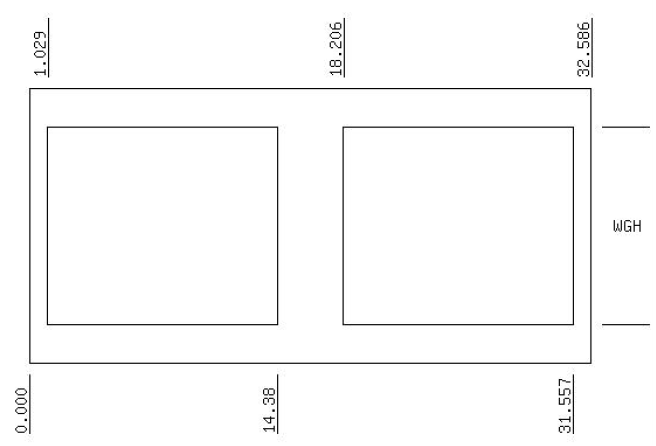}
 \caption{~}
 	 \label{fig:ref_desig2}
  \end{subfigure}%
 ~~~
  \begin{subfigure}[b]{0.47\textwidth}
	 \includegraphics[width=\textwidth]{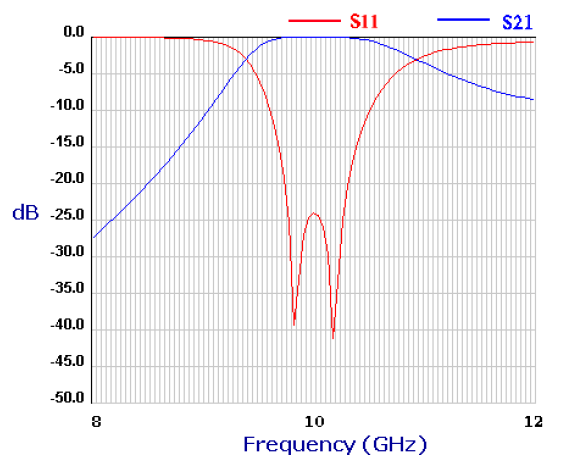}
 \caption{~}
	 \label{fig:ref_design2}
  \end{subfigure}%
\caption{Screen shots from a 2-pole metal insert waveguide filter design obtained through \texttt{www.guidedwavetech.com} in 2015 (the tool is now taken off line). (a) Dimensions in mm of metallic structure to be inserted in the waveguide. Only the part contained within WGH (set to $10\unit{mm}$) is relevant. (b) filter response of the design. Since the company lives off selling their software, we find the example meets today's industry standards albeit real applications normally require more complex characteristics, but may be based on the same design principle.}
  \label{fig:ref_design}
\end{figure}

The performance of a waveguide is often measure by its scattering, or $S$-parameters which provide information on the reflection and transmission of a wave in a waveguide. The $S$-parameters for a two port system can be computed as follows when only port 1 includes excitation:
\begin{align}
S_{11}(\omega) &= \frac{\int_{\Gamma_1} (\bm{E}-\bm{E}_{\text{inc}}) \cdot \bar{\bm{E}}_{\text{inc}} \text{d}\Gamma}
{\int_{\Gamma_1} |\bm{E}_{\text{inc}}|^2\text{d}\Gamma},\\
S_{21}(\omega) & = \frac{\int_{\Gamma_2} \bm{E} \cdot \bar{\bm{E}}_{\text{inc}} \text{d}\Gamma}
{\int_{\Gamma_2} |\bm{E}_{\text{inc}}|^2\text{d}\Gamma},
\end{align}
where $\bar{(\cdot)}$ refers to complex conjugate and $S_{11}$ corresponds to the reflection at surface $\Gamma_1$ while $S_{21}$ corresponds to the transmission at  $\Gamma_2$. A possible objective -- or fitness -- function for a given filter can be therefore be specified as a function of the $S$-parameters over some frequency range. To illustrate how this is to be done we use the pass band filter shown in Fig. \ref{fig:ref_design}. From the frequency sweep in Fig. \ref{fig:ref_design2} it is clear that for frequencies lower than 9.5GHz and above 10.5GHz, the transmission should be zero (or very small), 
while the transmission should be full, i.e. $S_{21}=1$, for the range between 9.5GHz and 10.5GHz. We can state this formally in terms of a discrete frequency list and two index sets
\begin{align}
f & = \{ f_1^z = 8,f^z_2 = 9, f_3^f = 9.5,
  			f_4^f = 10,f_5^f = 10.5,f_6^z = 11, f_7^z = 12 \} \unit{GHz}, \\ 
\mathbb{I}_f &= \{3,4,5\},\\
 \mathbb{I}_z &= \{1,2,6,7\},
\end{align}
where $f_i^f$, $\mathbb{I}_f$ and $f_i^z$, $\mathbb{I}_z$ refers to full and zero transmission respectively. The discrete frequency list can of course be 
refined or coarsened to fit a specific design task. The objective function can now be stated in terms of the transmission parameter as the following non-smooth maximization problem
\begin{equation}
\label{eq:objective_ref}
\Phi =   \max[ ~\min\limits_{j \in \mathbb{I}_f } \{ |S_{21}(\omega_j)| \} , ~\min\limits_{i \in \mathbb{I}_z } \{ 1 - |S_{21}(\omega_i)| \} ~],
\end{equation}
It is clear that $\Phi$ takes values between 0 and 1 and that $\Phi=1$ corresponds to the optimum. The non-smoothness can be remedied by introducing a bound formulation \citep{bendsoe2003}. The bound formulation can in turn be stated in several different ways, which will be discussed in the following section. 
%

%

%
%
\section{Two phase optimization methodology}
%
%
Even if a microwave filter is not based on cascading filter elements, it is unlikely to obtain good filter characteristics without a multipole design. 
From experience we have not been able to generate multipole design starting from a non-resonating starting guess since the optimization tends to get 
stuck in a local minimum containing only one pole. In recognition of this issue and the fact that we require our design methodology to be able to start from a uniform initial design, we have divided the optimization procedure in two phases: Phase 1 consists of 
designing a set of smaller resonators which can be used as initial guess for the full filter optimization, thus (over)populating the initial guess with several resonances. Phase 2 then consists of designing one integrated multipole structure that make up the overall filter response we are seeking.

The Phase 1 design problem(s), i.e. the resonator design, can be done in several ways using topology optimization. One way would 
be to do an eigenvalue target optimization where the difference between the resonance frequency and a target frequency is optimized 
(see e.g. \citet{Jensen2006}). This approach requires the solution of a quadratic eigenvalue problem due to the system damping, i.e. 
conductivity, which is currently not available within the used numerical framework. We therefore choose another design methodology capable of obtaining resonator 
designs which is based on the $S$-parameters and the non-smooth objective in Eq.  \eqref{eq:objective_ref}. Restating this as a bound formulation we obtain the following smooth optimization problem
\begin{equation}
\label{eq:opt_prob_p1}
\begin{array}{clll}
  \max \limits_{\beta \in \mathbb{Re}^+, \rho \in \mathbb{Re}^n} & \Psi_1 = \beta, &  \hspace{1cm}& \text{objective function} \\[3mm]
\text{subject to}  & \beta - |S_{21}(\omega_i)| \le 0, & i \in \mathbb{I}_f, & \text{full transmittance}\\[3mm]
~ & \beta - 1 + |S_{21}(\omega_k)| \le 0, & k \in \mathbb{I}_z, &  \text{zero transmittance}\\[3mm]
~ & \bm{S}(\omega_j) \bm{E}-\bm{f}(\omega_j) = 0,  & j \in \mathbb{I}_f \cup \mathbb{I}_z, & \text{state equation}\\[3mm]
  ~  & 0 \le \rho^e \le 1, & e = 1,n .& \text{design variable bounds}
  \end{array}
\end{equation}
Note that the extra variable $\beta$ is required to transform the objective function from Eq. \eqref{eq:objective_ref} into a smooth optimization problem that is tractable by standard gradient methods. 
Dependent on the specific filter design task, the number of needed Phase 1 problems will change and note that all Phase 1 problems are completely decoupled and can thus be solved completely independent of each other, i.e. they are embarrassingly parallel. 
It is an important observation that the formulation in Eq. \eqref{eq:opt_prob_p1} puts equal weight on transmission and reflection. That is, the optimizer will first try to reach a 
point in which all $S_{21}$ parameters equals $0.5$ and then subsequently expand the difference from the center point. 

The Phase 2 design problem is to obtain the full filter functionality. The starting point for Phase 2 are the results of Phase 1, however, the full filter design problem requires a different objective function. That is, since $\Psi_1$ puts an equal weight on transmission and reflection it is therefore likely to lead to designs with $S_{21}$ less than one and $S_{11}$ larger than zero. For the Phase 2 objective function the idea is to decouple transmission and the reflection. This is accomplished by introducing yet a bound variable $\kappa$ that only affects the reflection whereas $\beta$ only affects the transmission. 
Finally, it is desired that the Phase 2 objective function favors full transmission to that of full reflections. This means that a simple difference between $\kappa$ and $\beta$ does not suffice. However, from band gap optimization \citep{Sigmund2003a} it is known that a relative gap measure works very well in practice which leads to the following problem statement for Phase 2:
\begin{equation}
\label{eq:opt_prob_p2}
\begin{array}{clll}
  \max \limits_{\beta \in \mathbb{Re}^+, \kappa \in \mathbb{Re}^+, \rho \in \mathbb{Re}^n} &  \Psi_2 = \frac{\beta-\kappa}{\beta+\kappa+1} + \frac{1}{2},&  \hspace{1cm}& \text{objective function} \\[3mm]
\text{subject to}  & \beta - |S_{21}(\omega_i)| \le 0, & i \in \mathbb{I}_f,  & \text{full transmittance}\\[3mm]
~ & |S_{21}(\omega_k)| - \kappa \le 0, & k \in \mathbb{I}_z, & \text{zero transmittance}\\[3mm]
~ & \bm{S}(\omega_j) \bm{E}-\bm{f}(\omega_j) = 0,  & j \in \mathbb{I}_f \cup \mathbb{I}_z,  & \text{state equation}\\[3mm]
  ~  & 0 \le \rho^e \le 1, & e = 1,n. & \text{design variable bounds}
  \end{array}
\end{equation}
The addition of one to the denominator in $\Psi_2$ is needed to avoid zero sensitivities in the cases where either $\beta$ or $\kappa$ is zero. This is similar to the argument used for deriving the RAMP interpolation scheme as presented in \citet{Stolpe2001}. The $1/2$ is added to the Phase 2 objective for consistency such that both $\Psi_1$ and $\Psi_2$ take values between 0 and 1, where 1 refers to optimal performance of the system.

Both the Phase 1 and 2 optimization problems are solved using a gradient based optimization method with sensitivities obtained from the standard adjoint method, see e.g. \citet{Christensen2009}. The optimization algorithm used is the Method of Moving Asymptotes (MMA) \citep{Svanberg1987,Aage2013b}. 
\section{Design filtering and continuation approach}
Solving metallic optimization problems involving wave propagation poses a difficult task in terms of high contrast i.e conductor to air/dielectric \citep{Aage2010,Hassan2013}, and many local minima \citep{Diaz2010}. The used design interpolation and the two phase optimization setup alleviates these issues partially but to make the procedure robust we also include image processing filtering in conjunction with a continuation approach. We apply a standard convolution type filter to the design problem in terms of a Poisson type equation, i.e
\begin{equation}
−r^2 \nabla^2 \tilde{\rho} + \tilde{\rho} =  \rho
\end{equation}
where $\tilde{\rho}$ is the filtered design field and $r$ corresponds to the radius of influence for the filter. The filter problem is solved with pure Neumann conditions by a linear finite element model and included in the sensitivity calculation by the chain rule cf. \citep{Lazarov2010f}.

Image filtering naturally introduces gray-scale, i.e. blurred interfaces, in the physical design field. This is undesirable for several reasons. Firstly, graded materials are expensive to manufacture (if possible) and secondly, resonant structures, such as metallic filters, obtain much of their performance from sharp material interphases. 
Therefore we suggest the following continuation approach for the length scale parameter $r$. The layout of the scheme is constructed based on our numerical experiments similar to the works of e.g \citet{Alexandersen2015a} and \citet{Wang2011a}. For the dimensions used in the reference design of Fig.  \ref{fig:reference_design_optimization} we have settled on the following filter values and update intervals
\begin{equation}
r \in \{ 3.0,1.5,0.75,0.3,0.125\}\text{mm}
\end{equation}
The filter radius is updated at every 30th design cycles unless the optimization process converges due to other reasons, i.e. reaching a stationary point. We remark that the optimization process is restarted when the filter radius is updated to acknowledge the non-smooth nature of a sharp parameter change.

\section{Results}
In this section we first verify the method and validity of the chosen reference example. Subsequently it is showed how sharper filter characteristics can be obtained on the same design space with more design freedom given.

\subsection{Model validation and reference design}

\begin{figure}
	\centering
	\includegraphics[width=\textwidth]{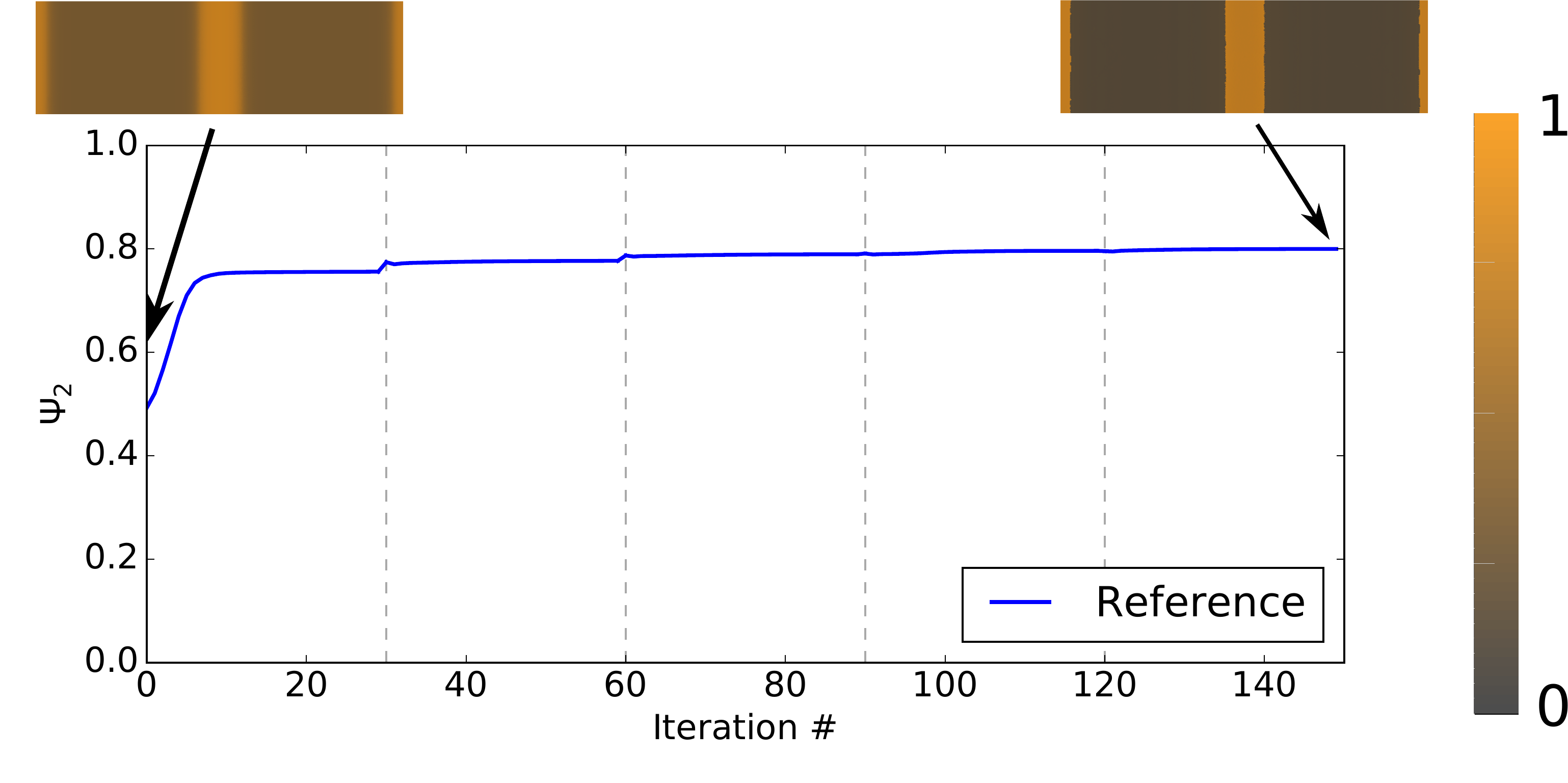}
	\caption{Iteration history including initial guess and final design for the filter which will be used as a reference. The color indicates design interpolation value of $\rho$. The final filter response can be found in Fig. \ref{fig:s_param}(b).}
	\label{fig:reference_design_optimization}
\end{figure}

For model validation and later benchmarking, a classical 2-pole insert filter design is chosen with geometry and characteristics as depicted in Fig. \ref{fig:ref_design}. The filter 
design method is still in use today, and we therefore consider it a realistic application example.
Formulating the filter characteristics in terms of the optimization problem in Eq. \eqref{eq:opt_prob_p2} with transmission from $9.8\unit{GHz}$ to $10.4\unit{GHz}$, we have
\begin{align}
\nonumber
f &= \{ f^z_1 = 8,\, f^z_2 = 8.5,\, f^z_3 = 9.0,\, 
						f^f_4=9.8,\, f^f_5=10.1,\, f^f_6=10.4,\\
			&\qquad \qquad 	f^z_7=11.2,\, f^z_8 = 11.6,\, f^z_9 = 12.0\} \unit{GHz}, \\
\mathbb I _f &= \{4,5,6\}, \\
\mathbb I_z &= \{1,2,3,7,8,9\}.
\end{align}
To verify that the reference filter indeed is a locally optimal solution 
this design is used as a start guess for the optimization with the only modification that in order to ensure good design gradients, the void 
regions were replaced with intermediate material ($\rho=0.3$). The results of the optimization is seen in Fig. \ref{fig:reference_design_optimization}, where it is 
seen that no substantial design changes appeared. This is further confirmed by comparing Fig. \ref{fig:ref_design2} to the filter characteristics in Fig. \ref{fig:s_param}(b-c) for the reference design from Fig.
\ref{fig:reference_design_optimization}. 

The optimization also validates the code. First of all, the implementation is verified by the fact that the results in Fig. \ref{fig:ref_design}(b) are reproduced. Furthermore, 
it confirms that the optimization procedure is correctly implemented due to the monotonous behavior of the objective function with respect to iteration number (except for when filter 
radius is changed due to the continuation approach). The optimization result also confirms that the design we want to use as a benchmark is a local minimum with respect to the desired objective function.

Finally note that the reference filter insert design is \emph{not} designed for a Printed Circuit Board (PCB), i.e. where a metal structure is fabricated on a dielectric background using photo lithographic etching. 
Using PCBs provides full 2D topological freedom and is included in our numerical model and design methodology. In the following we therefore assume that the structures can be attached to a dielectric with 
the same permittivity and permeability as air such that direct comparison with the reference design is possible.


\subsection{Topology optimized filter design}
Using the validated code and the reference example as a benchmark, we now show how steeper filter characteristics can be obtained through topology optimization by allowing complete 2D design freedom. 
The procedure takes outset in a uniform start guess, and is as such not seeded towards any specific structures to begin with.

\subsubsection{Phase 1 -- single resonator optimization}
As described, the first task is to design a number of independent resonators using smaller design volumes. For the pass band filter studied here we choose to start out with three initial resonators. The total design domain has a length of $32.6\unit{mm}$ equivalent 
to  that of the reference design. We include an air gap of $1\unit{mm}$ between the each of the three resonators to ensure they do not short-circuit when cascaded, which means that each resonator has a design space of $10.2\unit{mm}$. 
As target frequencies for each of the Phase 1 resonators we choose three separate resonances inside the passband given by the center frequencies below
\begin{align}
f^1_c = 9.8 \unit{GHz}, \quad f^2_c = 10.1 \unit{GHz}, \quad f^3_c = 10.4 \unit{GHz}.
\end{align}
The set of frequencies to maximize and minimize in Eq. \eqref{eq:opt_prob_p1} for the three different problems are given as:
\begin{align}
\nonumber
f &= f_c \times \{ f^z_1 = 1-3b,\, f^z_2 =  1-2b,\, f^z_3 1-b,\, f^f_4=1,\\
	&\qquad \qquad f^z_5= 1+b,\, f^z_6=1+2b,f^z_7 =\,1+3b \} , \\
	\mathbb I_f &= \{4\}, \\
	\mathbb I_z &= \{1,2,3,5,6,7\},
\end{align}
where $f_c = f_c^1,\, f_c^2,\,f_c^3$ corresponds to each of the three resonator design respectively, and $b=0.08$ is a parameter that defines the trade off between transmission and reflection. 
We find that giving the reflection frequencies in terms of a fractional bandwidth in general gives better designs than when specifying equidistant frequencies. This is simply explained by the fact that a $100\unit{Mhz}$ difference at two different frequencies does not translate to the same sharpness of a resonator.

\begin{figure}
\centering
\includegraphics[width=\textwidth]{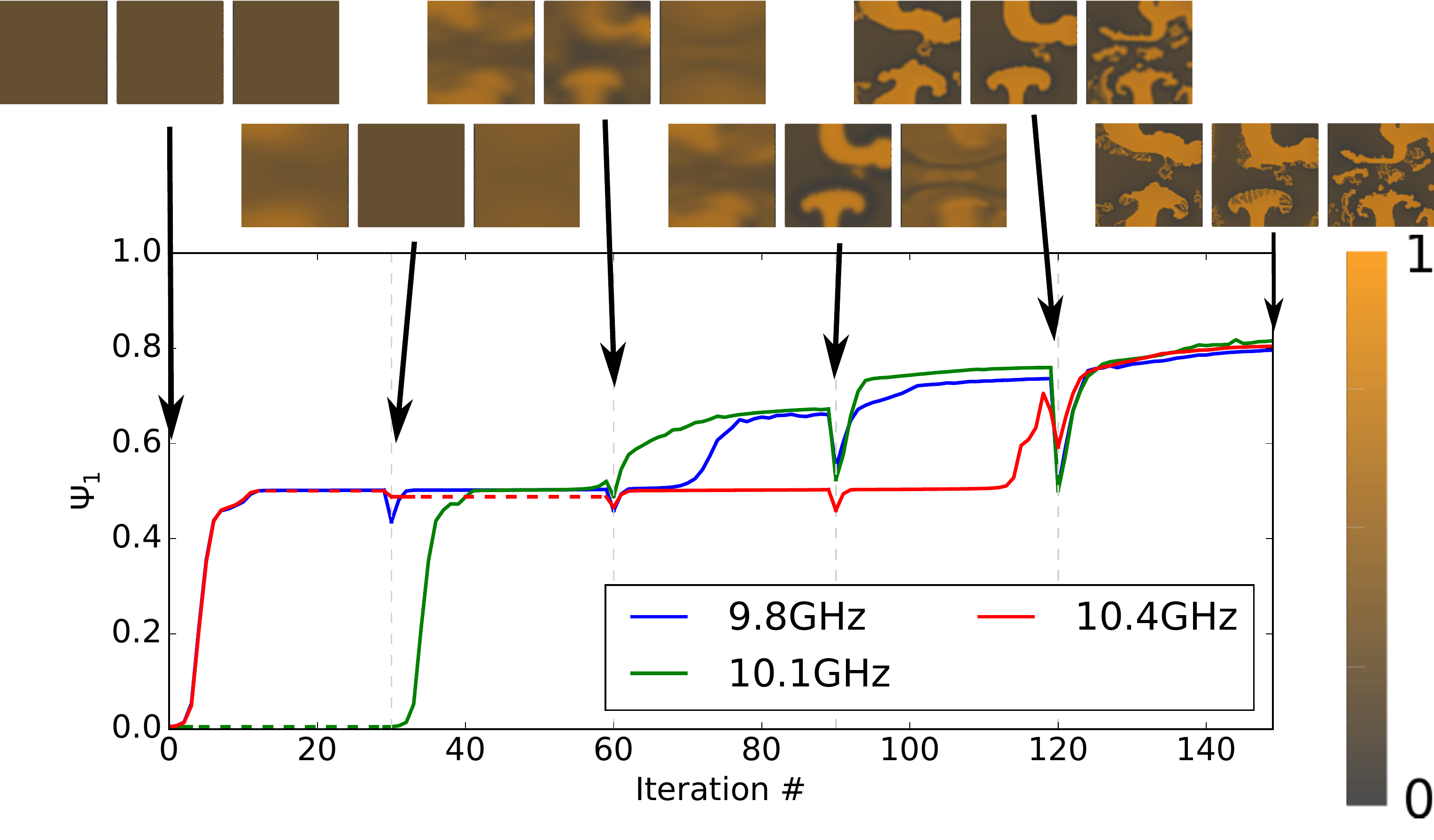}
\caption{Overview of Phase 1 of the optimization process for the three small, individually optimized, resonators that will make out the start guess for Phase 2. The three intermediate designs are shown above with rising design frequency from left to right. The filter size was decreased every 30th iteration, and the dashed lines indicate that optimization was stopped due to lack of changes in the design.}
\label{fig:P1result}
\end{figure}

The results of the design process and the intermediate steps are seen in Fig. \ref{fig:P1result}. Several trends can be noticed for all three optimizations (the trends are also observed in other design trials we have performed): 1) The optimization procedure spend many iterations trying to overcome the state where transmission equals reflection ($\Psi_1=0.5$), which we interpret as a difficulty of trying to create a structure which gaps reflection and transmission at the desired frequencies. Until then, the algorithm potentially decreases both transmission and reflection at desired frequencies. 
This is not necessarily a desired route to take for the algorithm, and we hope a future formulation can get rid of this in order to speed up the process and to obtain better resonators at this part of the design stage. 
2) The optimizations at different filter radii does not necessarily converge within the given 30 iterations, but we find that better designs are generally obtained by stopping the process prematurely. 
This is explained by the fact that wave propagation design problems are proned to many (strong) local minima  \citep{Aage2011}, and hence that we wish to terminate the optimization process prematurely before the continuation strategy has reached its end.
3) Topological changes happen several times during the optimization which is a sign of a good degree of design freedom, especially considering how easily metallic microwave problems may get stuck in local minima. 

In the end, three soft resonators are found of more or less similar quality based on the objective value. Their filter response is seen in Fig. \ref{fig:s_param}(a).

\subsubsection{Phase 2 -- complete filter optimization}
\begin{figure}
	\centering
	\includegraphics[width=\textwidth]{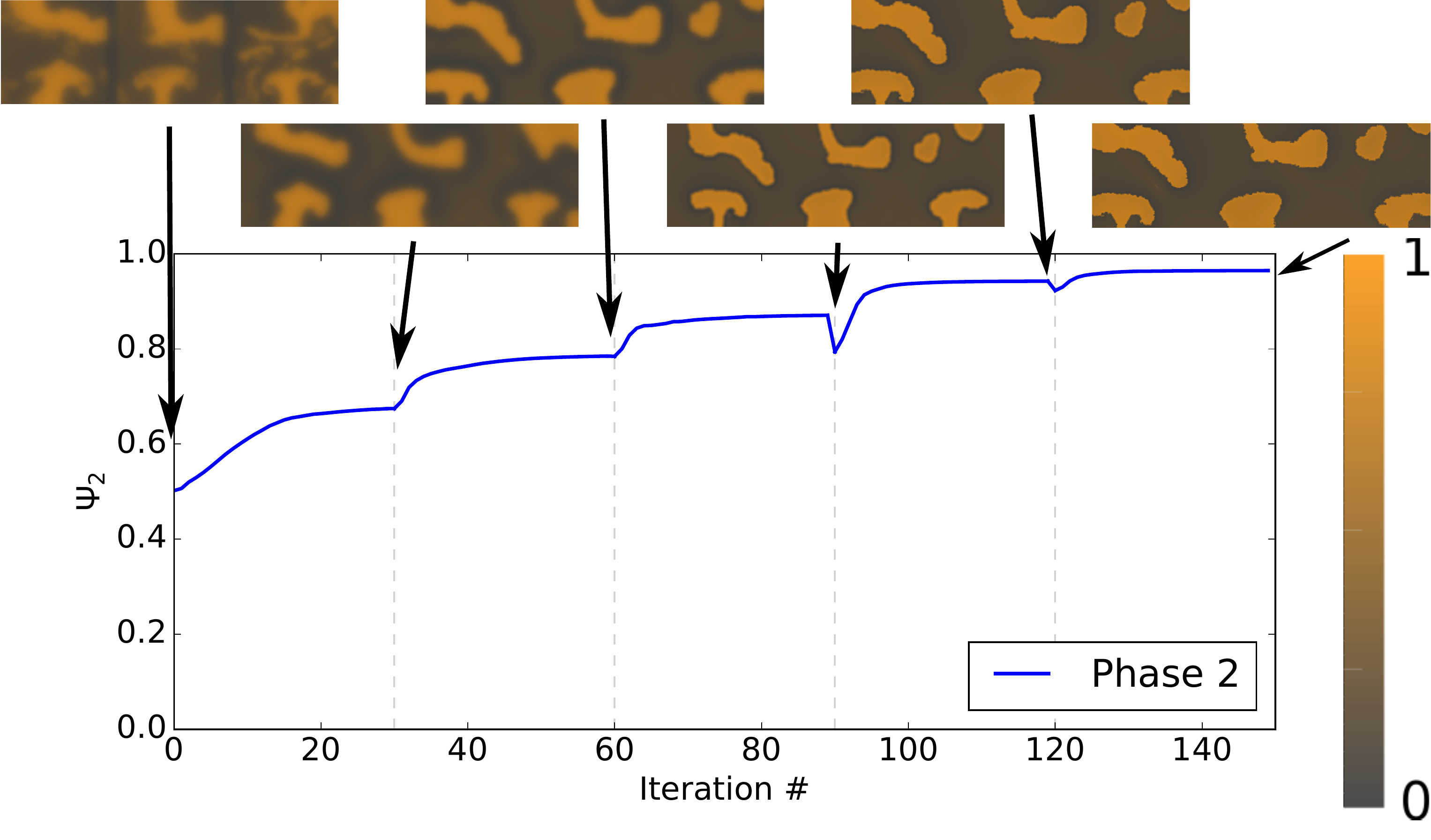}
	\caption{Overview of Phase 2 of the optimization process.}
	\label{fig:P2result}
\end{figure}

The three single resonators are now placed next to each other -- with a spacing of $1\unit{mm}$ -- and used as  starting 
guess for the actual filter characteristics optimization process, i.e. the Phase 2 problem in \eqref{eq:opt_prob_p2}. Just as for the reference design, the frequencies for the optimization are
\begin{align}
\nonumber
f &= \{ f^z_1 = 8,\, f^z_2 = 8.5,\, f^z_3 = 9.0,\, 
						f^f_4=9.8,\, f^f_5=10.1,\, f^f_6=10.4,\\
			&\qquad \qquad 	f^z_7=11.2,\, f^z_8 = 11.6,\, f^z_9 = 12.0\} \unit{GHz}, \\
\mathbb I _f &= \{4,5,6\}, \\
\mathbb I_z &= \{1,2,3,7,8,9\}.
\end{align}
The objective value at first iteration is $\Psi_2 = 0.502$, which means that an ever so slight gap between transmission and rejection is present, which is crucial for the optimization 
since we find that the formulation tends to get stuck at $\Psi_2 = 0.5$ as mentioned earlier. However, the Phase 2 objective in Eq. \eqref{eq:opt_prob_p2} tends to perform significantly 
better when past this critical point. The outcome of the design procedure is seen in Fig. \ref{fig:P2result}.

\begin{figure}
	\centering
	\begin{tabular}{ccl}
		\includegraphics[width=0.45\textwidth]{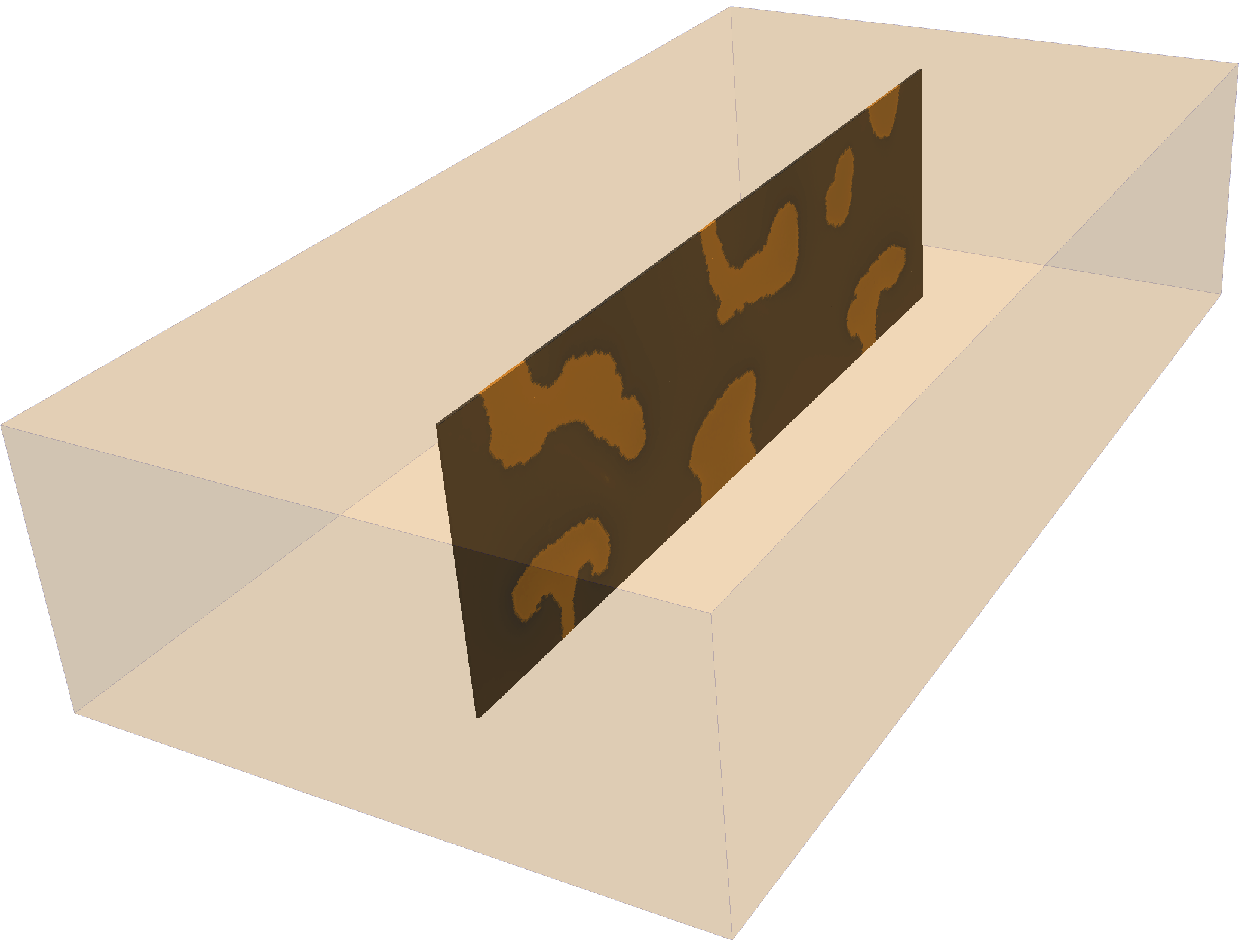} &
		\includegraphics[width=0.45\textwidth]{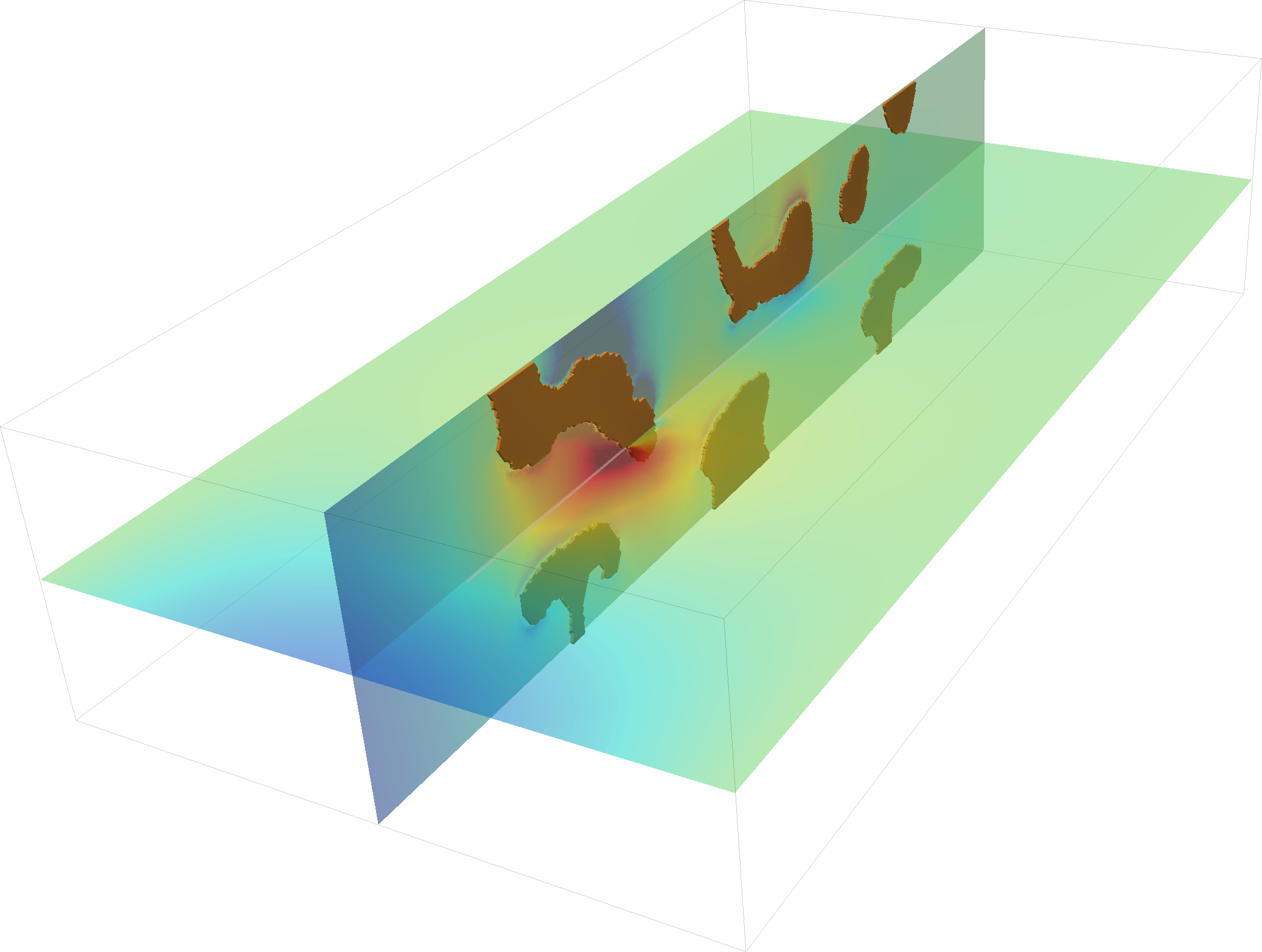} \\
		(a) Final design & (b) $8 \unit{GHz}$\\
		\includegraphics[width=0.45\textwidth]{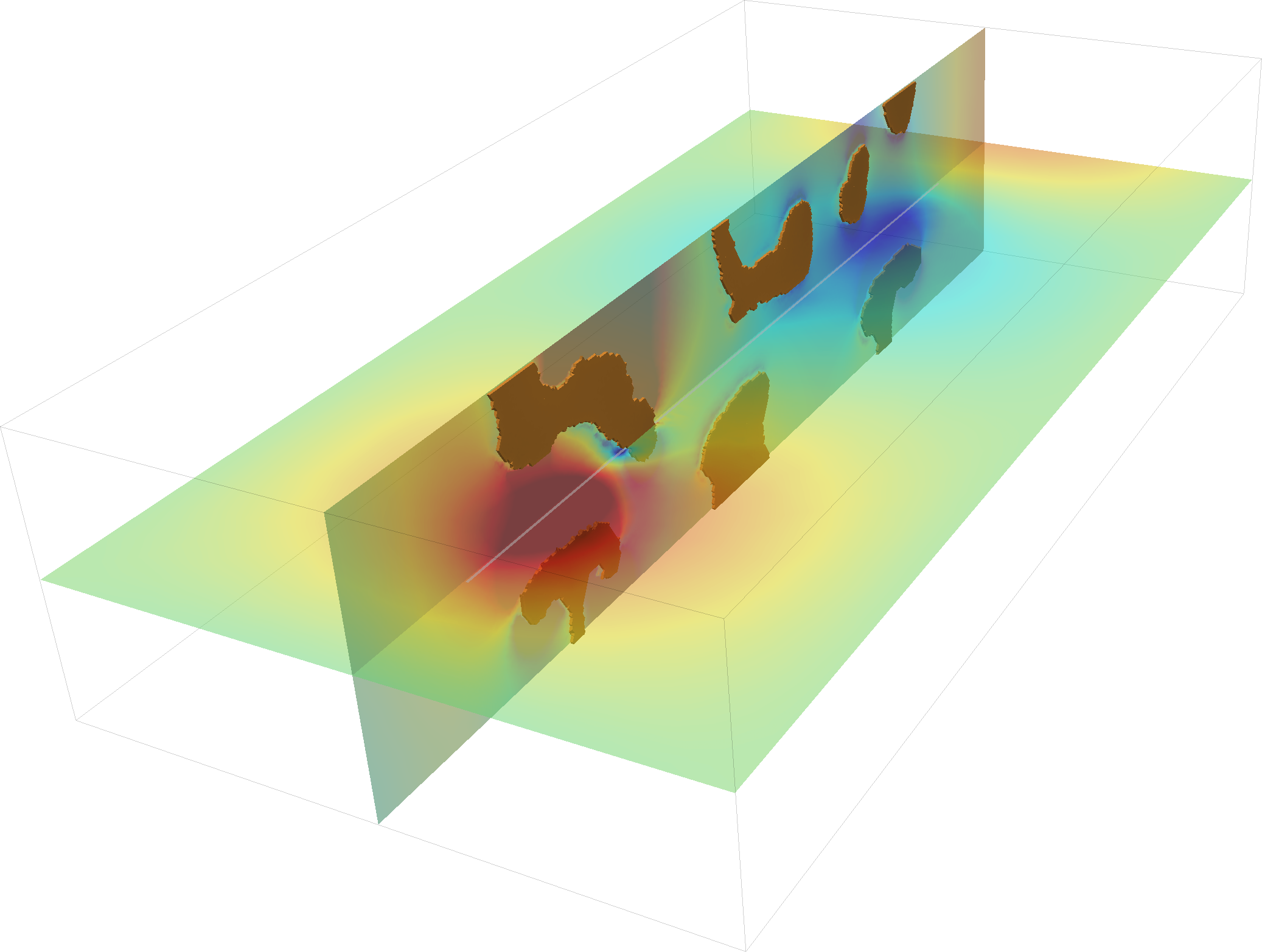} &
		\includegraphics[width=0.45\textwidth]{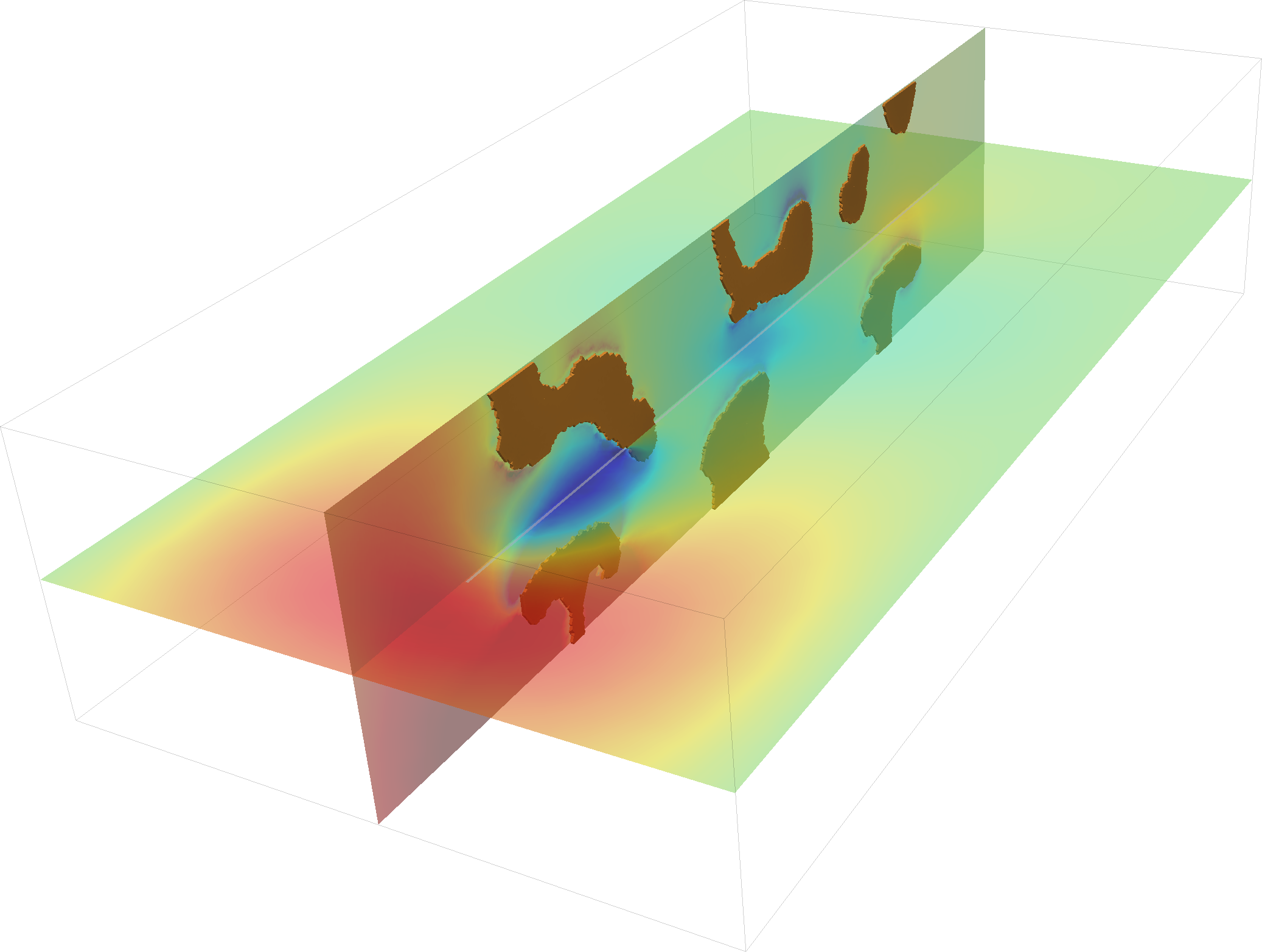} &
		\includegraphics[width=0.1\textwidth]{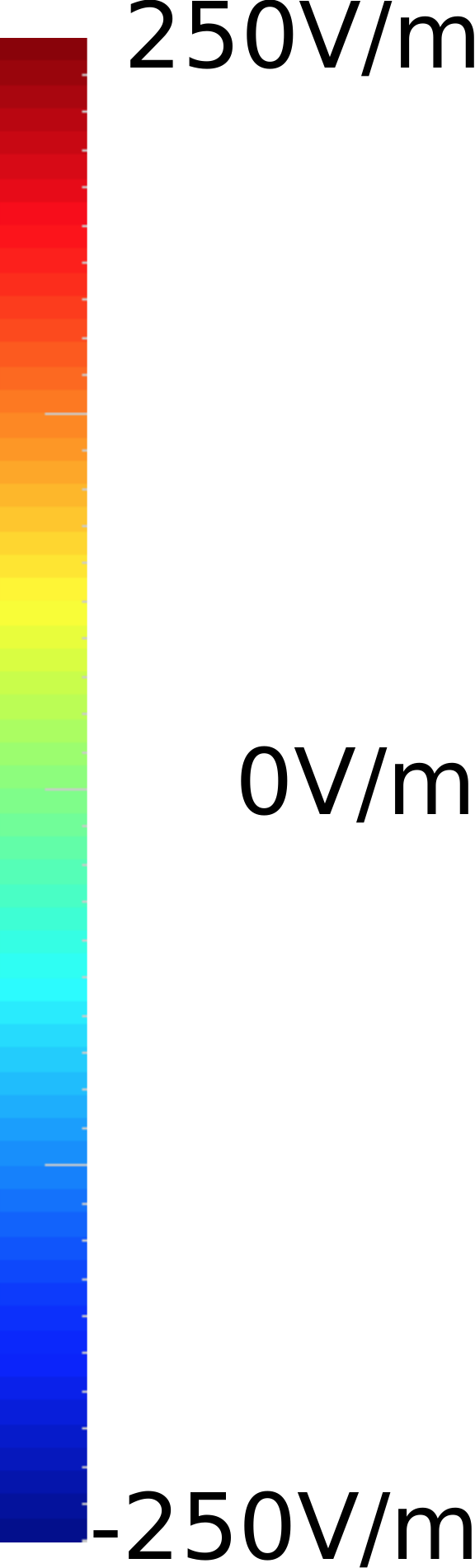}
		\\ 
		(c) $10.1 \unit{GHz}$ & (d) $12 \unit{GHz}$
	\end{tabular}
	\caption{(a) Visualization of how the final design would be sitting in a waveguide in order to act as a filter. The structure is a dielectric material 
	with copper imprint (PCB). The relative dielectric constant is set to 1, but could easily be chosen to fit standard PCB materials. (b-d) Electric field $y$-component 
	for $8 \unit{GHz}$, $10.1\unit{GHz}$ and $12\unit{GHz}$ respectively. Since the propagating mode only has a $y$-component, most of the energy is contained here. It is seen 
	how the wave mode is reconstructed at the other side for $10.1\unit{GHz}$, whereas it does not propagate through the filter at the higher/lower frequencies. The input power corresponds to $1\unit{W}$.}
	\label{fig:finalDesignLarge}
\end{figure}

During the first filter radius in the continuation process we observe that small features disappear and that larger features significantly change their shape. Furthermore, several 
changes in topology takes place during the second step in the continuation process. This clearly indicates how the previously obtained resonators in themselves 
do not provide a good filter response but serve well as a template for creating a larger filter. After this, the topology seems locked and the remaining 
optimization process constitutes a a shape optimization procedure. The objective of the final design is $\Psi_2 = 0.964$ as compared to the reference design which 
has an objective of $\Psi_2 = 0.800$. The microwave filter, as it would be situated in an actual waveguide, is depicted in Fig. \ref{fig:finalDesignLarge}. Electrical field 
amplitudes at one passband and two stop band frequencies are also shown and clearly demonstrates the filter functionality.

\begin{figure}
	\centering
	\begin{tabular}{ccc}
		\includegraphics[width=0.5\textwidth]{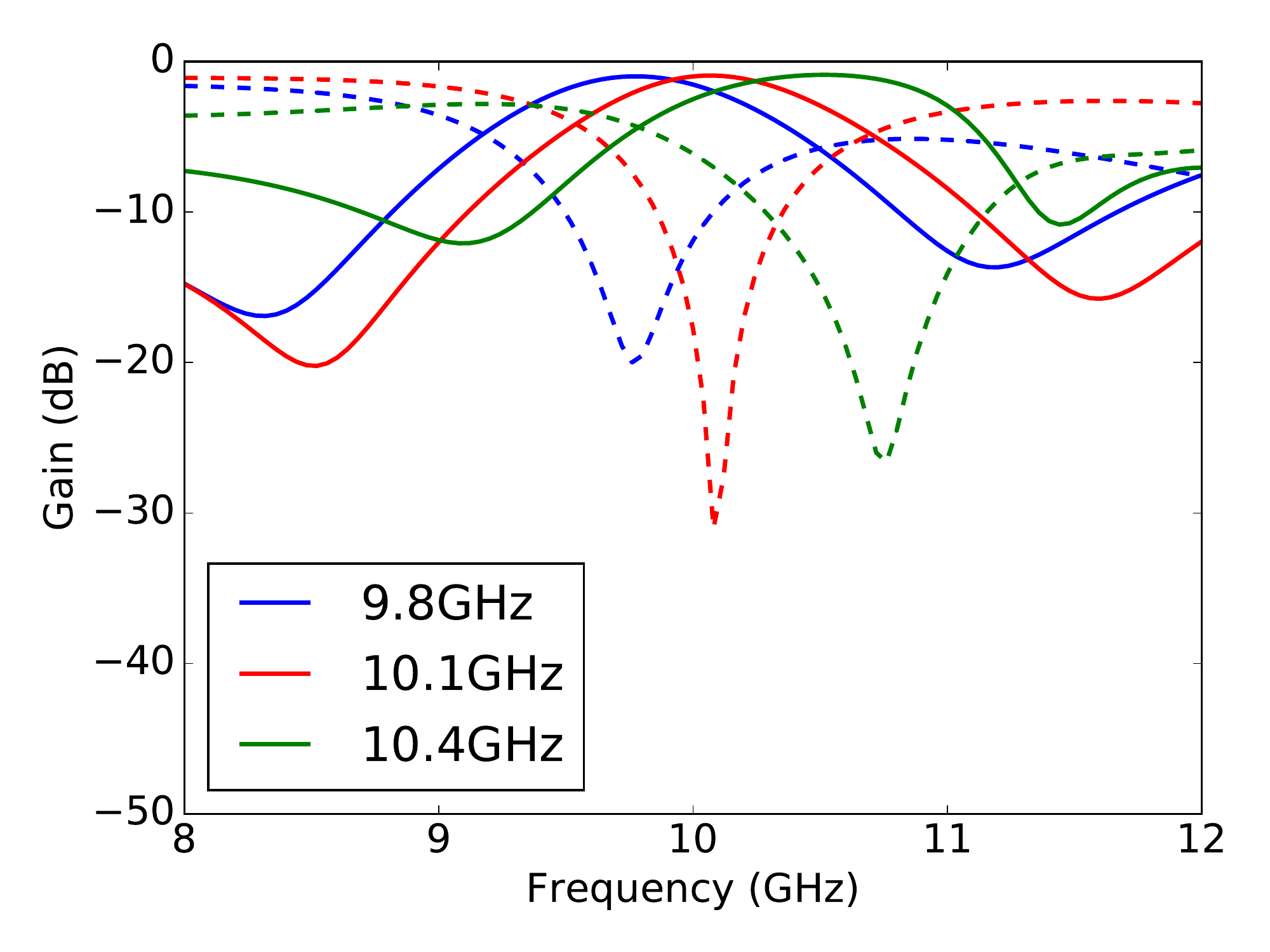} &
		\includegraphics[width=0.5\textwidth]{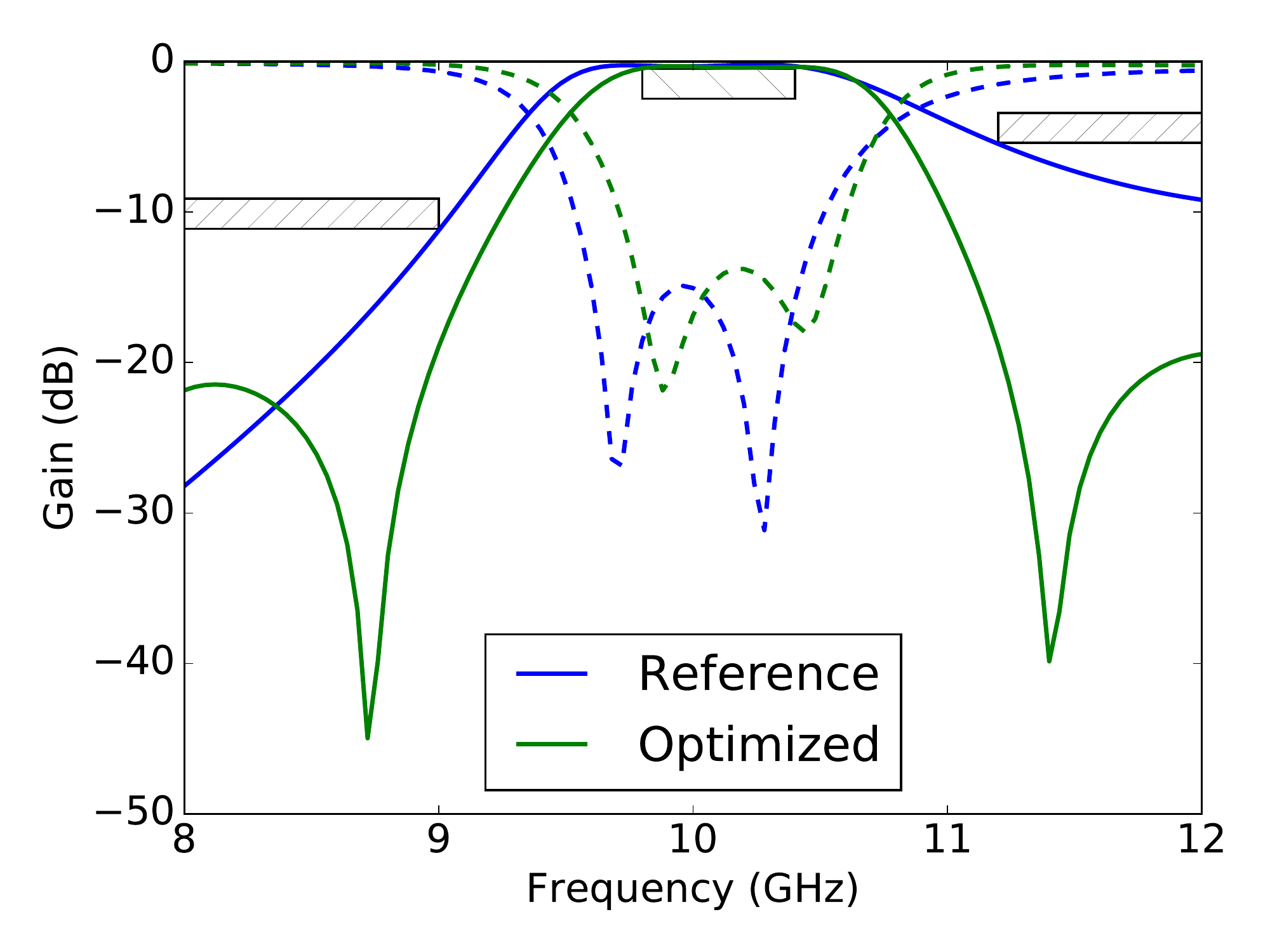} \\
		(a) & (b) \\
		\includegraphics[width=0.5\textwidth]{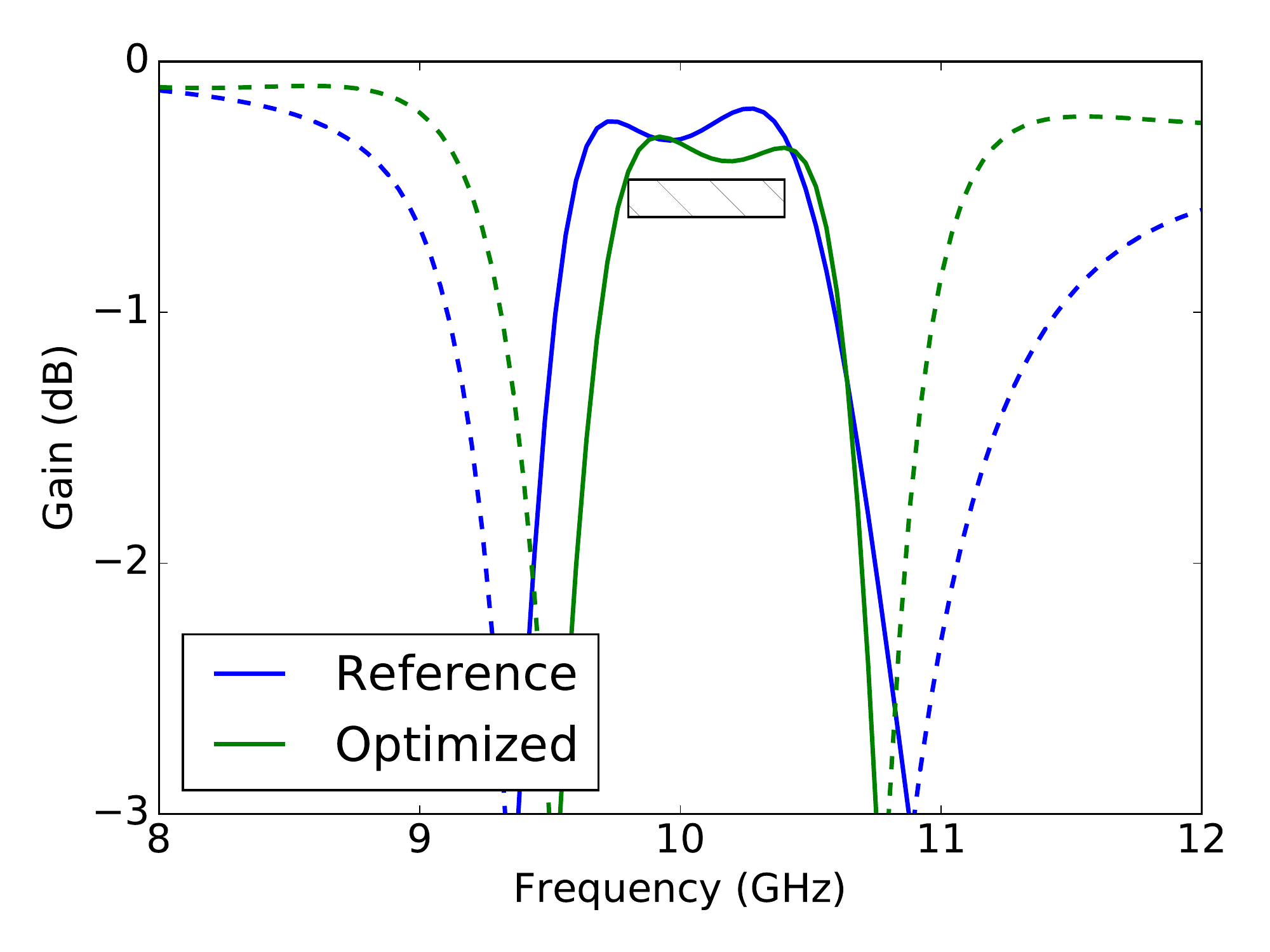} &
		\includegraphics[width=0.5\textwidth]{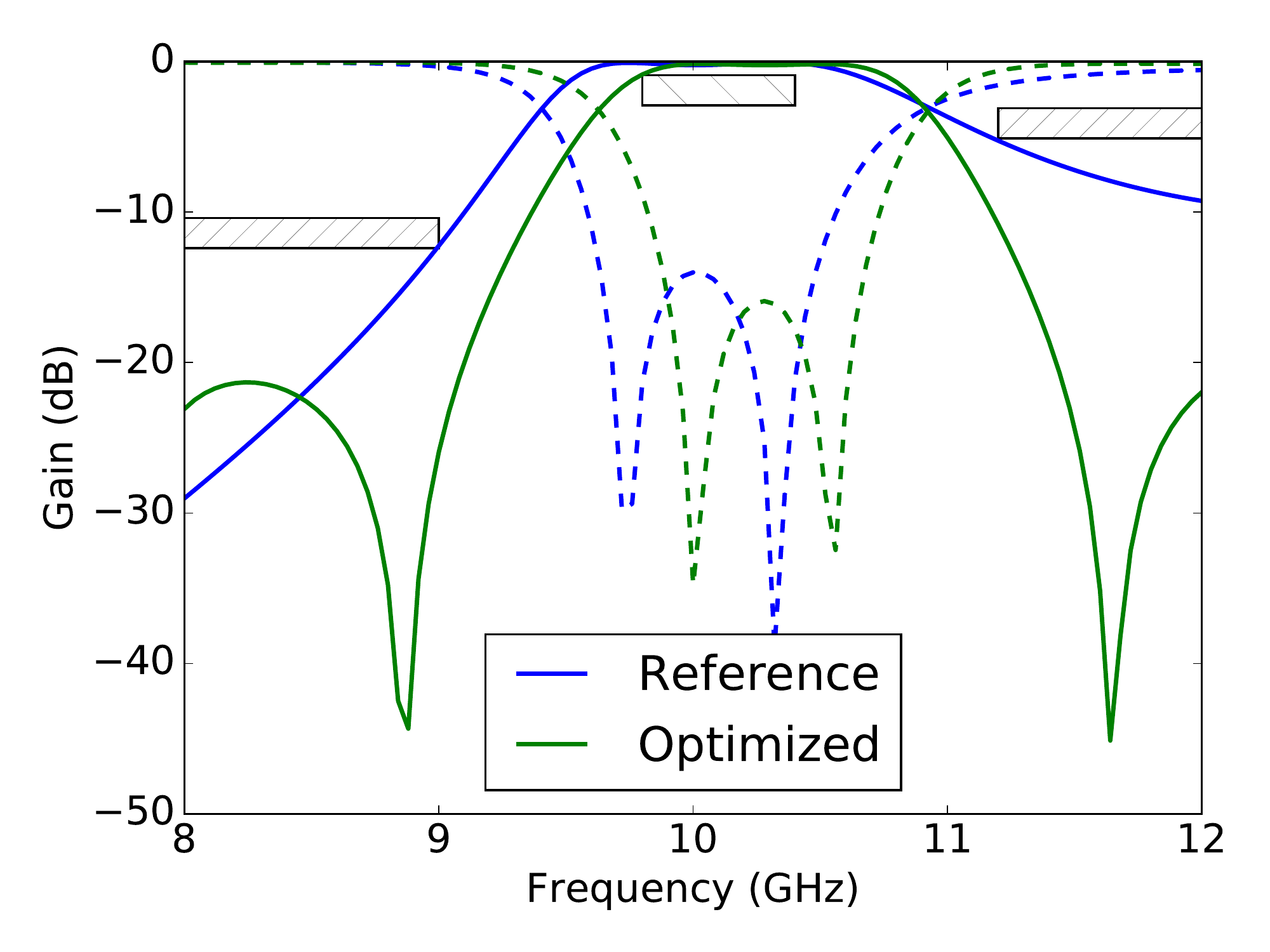} \\
		(c) & (d) 
	\end{tabular}
	\caption{Filter characteristics for the different optimization results. The solid lines are transmission ($S_{21}$) and the dashed lines are reflection ($S_{11}$). (a) Characteristics of the three filters designed in the Phase 1 optimization in Fig. \ref{fig:P1result}. (b) Filter characteristics of the optimized reference filter in Fig. \ref{fig:reference_design_optimization} and the freely optimized filter in Fig. \ref{fig:P2result}. (c) Close-up of result in (b). (d) Filter response of a post-processed version of the filters where the design elements have been exchanged with PEC. All hatched areas indicates the ranges of frequencies defined for either transmission or rejection in the optimization problem.}
	\label{fig:s_param}
\end{figure}

A comparison of the final filter characteristics to the reference filter is seen in Fig. \ref{fig:s_param}(b-c). An insertion loss of $0.5\unit{dB}$ is seen 
for both filters. The filters are simulated using PEC in Fig. \ref{fig:s_param}(d) to show the limited influence of our upper bound for material interpolation.

In comparing the filters, the roll off for the optimized filter is much sharper than for the reference design. This seems to be due to the placement of two anti-resonances/transmission dips at 
roughly $8.75\,\mathrm{GHz}$ and $11.75 \,\mathrm{GHz}$. Such two resonances are not seen in the reference design, and they cannot easily be squeezed in 
using the current design approach without enlarging the design domain. In the transmission band, two poles are still present. One may have hoped for three poles in the transmission band, but in 
general we find that poles often merge during the optimization process. This is also why it is so important to start with a multipole design, since resonant structures 
contain strong local minima that are hard to get into as well as hard to get out of. For higher frequencies we see an improvement in the rejection band of $6\,\unit{dB}$ or more. For lower frequencies, 
we see a steeper transition for the optimized design than for the reference (due to the "anti-pole"), but for frequencies around $8\,\unit{GHz}$, the optimized design actually has a smaller 
rejection than the reference. This is attributed to the anti-resonance, but since our cost function looked at overall rejection and this increase occurs where the rejection is already quite 
high, we do not see this as an issue.

\section{Outlook}
For future work, we believe that several steps can be taken in order to obtain better results and a wider usability of the method. First of all, as seen from the iteration 
history for the Phase 1 design, it is difficult for the algorithm to move past $\Psi_1 = 0.5$, and there is no reason in the first place for the algorithm to go through a design with 
equal transmission and rejection. It is probably hard to find better objective functions and a possible solution to overcome this challenge is to optimize the resonators based on an 
eigenfrequency problem instead. 

It is also worth testing whether or not the reciprocity ($S_{21}=S_{12},S_{11}=S_{22}$) of the problem can be exploited. By having an objective that is the average of two test cases 
with waves launched from either end of the waveguide, better sensitivities may be obtained over the whole design domain. The idea behind this approach is similar to how the problems 
in topology optimized antenna design are inverted by first considering radiation and then illumination \citep{Aage2011}.

For filter design with more strict control of the response, asymptotic wave expansion (AWE) would help increasing solution performance and computational effort by providing a way to 
obtain a continuous filter response curve instead of the relatively low set of discrete data points we are using at the moment \citep{Jensen2007}. AWE would furthermore allow more 
detailed control of the filter characteristics, but at the cost of more constraints and hence a harder optimization problem. 

\section{Conclusions}
A metallic waveguide filter with improved performance as compared to a benchmark example has been designed using a topology optimization approach. The approach is implemented 
using the finite element method (FEM) and is divided into two design phases in order to obtain the strong resonances required for obtaining a filter with high transmission and 
rejection characteristics. The obtained design does not bear resemblance with any earlier design and thus indicates the potential for rethinking traditional filter design 
approaches for microwave engineering. We believe that this proof-of-concept is the first step towards having topology optimized microwave filters in engineering applications.


\section{Acknowledgements}
This work was funded by the Villum Foundation through the \textit{NextTop} project and the Danish National Advanced Technology Foundation through the grant \textit{Wireless Coupling in Small Autonomous Apparatus} (www.hoejteknologifonden.dk). The authors would like thank to Vitaliy Zhurbenko and the DTU TopOpt group for valuable insight and many fruitful discussion.
\bibliographystyle{abbrvnat}

\bibliography{./library}

\begin{thebibliography}{32}
\providecommand{\natexlab}[1]{#1}
\providecommand{\url}[1]{\texttt{#1}}
\expandafter\ifx\csname urlstyle\endcsname\relax
  \providecommand{\doi}[1]{doi: #1}\else
  \providecommand{\doi}{doi: \begingroup \urlstyle{rm}\Url}\fi

\bibitem[Aage(2011)]{Aage2011}
N.~Aage.
\newblock \emph{{Topology optimization of radio frequency and microwave
  structures}}.
\newblock Ph.d. dissertation, Techanical University of Denmark, 2011.

\bibitem[Aage and Lazarov(2013)]{Aage2013b}
N.~Aage and B.~S. Lazarov.
\newblock {Parallel framework for topology optimization using the method of
  moving asymptotes}.
\newblock \emph{Structural and Multidisciplinary Optimization}, 47\penalty0
  (4):\penalty0 493--505, 2013.
\newblock ISSN 1615147X.

\bibitem[Aage et~al.(2010)Aage, Mortensen, and Sigmund]{Aage2010}
N.~Aage, N.~A. Mortensen, and O.~Sigmund.
\newblock {Topology optimization of metallic devices for microwave
  applications}.
\newblock \emph{International Journal for Numerical Methods in Engineering},
  83\penalty0 (2):\penalty0 228--248, 2010.

\bibitem[Alexandersen et~al.(2016)Alexandersen, Sigmund, and
  Aage]{Alexandersen2015a}
J.~Alexandersen, O.~Sigmund, and N.~Aage.
\newblock {Large scale three-dimensional topology optimisation of heat sinks
  cooled by natural convection}.
\newblock \emph{International Journal of Heat and Mass Transfer}, 100\penalty0
  (SEPTEMBER):\penalty0 876--891, 2016.
\newblock ISSN 00179310.
\newblock \doi{10.1016/j.ijheatmasstransfer.2016.05.013}.
\newblock URL \url{http://dx.doi.org/10.1016/j.ijheatmasstransfer.2016.05.013}.

\bibitem[Amestoy et~al.(2000)Amestoy, Duff, and L'Excellent]{Amestoy2000}
P.~Amestoy, I.~Duff, and J.~L'Excellent.
\newblock {Multifrontal parallel distributed symmetric and unsymmetric
  solvers}.
\newblock \emph{Computer Methods in Applied Mechanics and Engineering},
  184\penalty0 (2-4):\penalty0 501--520, 2000.

\bibitem[Assadihaghi et~al.(2006)Assadihaghi, Bila, Durousseau, Baillargeat,
  Aubourg, Verdeyme, Rochette, Puech, and Lapierre]{Assadihaghi2006}
A.~Assadihaghi, S.~Bila, C.~Durousseau, D.~Baillargeat, M.~Aubourg,
  S.~Verdeyme, M.~Rochette, J.~Puech, and L.~Lapierre.
\newblock {Design of Microwave Components using Topology Gradient
  Optimization}.
\newblock In \emph{Microwave Conference, 2006. 36th European}, pages 462--465,
  sep 2006.
\newblock \doi{10.1109/EUMC.2006.281392}.

\bibitem[Balanis(2012)]{balanis2012}
C.~A. Balanis.
\newblock \emph{{Advanced Engineering Electromagnetics}}.
\newblock CourseSmart Series. Wiley, 2nd edition, 2012.
\newblock ISBN 9780470589489.

\bibitem[Bendsoe and Sigmund(2003)]{bendsoe2003}
M.~P. Bendsoe and O.~Sigmund.
\newblock \emph{{Topology Optimization: Theory, Methods and Applications}}.
\newblock Engineering online library. Springer, 2003.
\newblock ISBN 9783540429920.
\newblock URL \url{http://books.google.dk/books?id=NGmtmMhVe2sC}.

\bibitem[Byun and Park(2007)]{Byun2007}
J.-K. Byun and I.-H. Park.
\newblock {Design of dielectric waveguide filters using topology optimization
  technique}.
\newblock \emph{IEEE transactions on magnetics}, 43\penalty0 (4):\penalty0
  1573--1576, 2007.
\newblock ISSN 0018-9464.
\newblock URL \url{http://cat.inist.fr/?aModele=afficheN{\&}cpsidt=18661172}.

\bibitem[Choi et~al.(2012)Choi, Kim, Lee, and Byun]{Choi2012}
N.-S. Choi, D.-H. Kim, H.-B. Lee, and J.-K. Byun.
\newblock {Topology Optimization of Dielectric Resonator in 3-D Waveguide
  Structure Considering Higher Mode Incidence}.
\newblock \emph{Magnetics, IEEE Transactions on}, 48\penalty0 (2):\penalty0
  559--562, feb 2012.
\newblock ISSN 0018-9464.
\newblock \doi{10.1109/TMAG.2011.2176109}.

\bibitem[Christensen and Klarbring(2009)]{Christensen2009}
P.~W. Christensen and A.~Klarbring.
\newblock \emph{{An Introduction to Structural Optimization}}, volume 153.
\newblock 2009.
\newblock ISBN 9781402086656.
\newblock URL \url{www.springer.com/series/6557}.

\bibitem[Delhote et~al.(2011)Delhote, Bila, Baillargeat, Chartier, and
  Verdeyme]{Delhote2011}
N.~Delhote, S.~Bila, D.~Baillargeat, T.~Chartier, and S.~Verdeyme.
\newblock {Advanced design and manufacturing of microwave components based on
  shape optimization and ceramic stereolithography process}.
\newblock \emph{2008 IEEE MTT-S International Microwave Workshop Series IMWS on
  Art of Miniaturizing RF and Microwave Passive Components - Proceeding}, pages
  15--18, 2011.
\newblock \doi{10.1109/IMWS.2008.4782250}.

\bibitem[Diaz and Sigmund(2010)]{Diaz2010}
A.~R. Diaz and O.~Sigmund.
\newblock {A topology optimization method for design of negative permeability
  metamaterials}.
\newblock \emph{Structural and Multidisciplinary Optimization}, 41:\penalty0
  163--177, 2010.
\newblock ISSN 1615147X.
\newblock \doi{10.1007/s00158-009-0416-y}.

\bibitem[Erentok and Sigmund(2011)]{Erentok2011}
A.~Erentok and O.~Sigmund.
\newblock {Topology optimization of sub-wavelength antennas}.
\newblock \emph{IEEE Transactions on Antennas and Propagation}, 59\penalty0
  (1):\penalty0 58--69, 2011.
\newblock ISSN 0018926X.
\newblock \doi{10.1109/TAP.2010.2090451}.

\bibitem[Hassan et~al.(2013)Hassan, Wadbro, and Berggren]{Hassan2013}
E.~Hassan, E.~Wadbro, and M.~Berggren.
\newblock {Topology Optimization of UWB Monopole Antennas}.
\newblock Technical report, 2013.

\bibitem[Hassan et~al.(2014)Hassan, Wadbro, and Berggren]{Hassan2014}
E.~Hassan, E.~Wadbro, and M.~Berggren.
\newblock {Topology optimization of metallic antennas}.
\newblock \emph{IEEE Transactions on Antennas and Propagation}, 62\penalty0
  (X):\penalty0 2488--2500, 2014.
\newblock ISSN 0018926X.
\newblock \doi{10.1109/TAP.2014.2309112}.

\bibitem[Hassan et~al.(2015)Hassan, Noreland, Augustine, Wadbro, and
  Berggren]{Hassan2015}
E.~Hassan, D.~Noreland, R.~Augustine, E.~Wadbro, and M.~Berggren.
\newblock {Topology Optimization of Planar Antennas for Wideband Near-Field
  Coupling}.
\newblock \emph{IEEE Transactions on Antennas and Propagation}, 63\penalty0
  (9):\penalty0 4208--4213, 2015.
\newblock ISSN 0018926X.
\newblock \doi{10.1109/TAP.2015.2449894}.

\bibitem[Jensen(2007)]{Jensen2007}
J.~S. Jensen.
\newblock {Topology optimization of dynamics problems with Pade approximants}.
\newblock \emph{Proceedings of the 2011 American Control Conference}, \penalty0
  (April):\penalty0 1605--1630, 2007.
\newblock ISSN 0743-1619.
\newblock \doi{10.1002/nme.2065}.

\bibitem[Jensen and Pedersen(2006)]{Jensen2006}
J.~S. Jensen and N.~L. Pedersen.
\newblock {On maximal eigenfrequency separation in two-material structures: The
  1D and 2D scalar cases}.
\newblock \emph{Journal of Sound and Vibration}, 289\penalty0 (4-5):\penalty0
  967--986, 2006.
\newblock ISSN 0022460X.
\newblock \doi{10.1016/j.jsv.2005.03.028}.

\bibitem[Jin(2002)]{jin2002}
J.~M. Jin.
\newblock \emph{{The Finite Element Method in Electromagnetics}}.
\newblock A Wiley-Interscience publication. Wiley, 2002.
\newblock ISBN 9780471438182.

\bibitem[Karypis and Kumar(1999)]{Karypis1999}
G.~Karypis and V.~Kumar.
\newblock {A Fast and Highly Quality Multilevel Scheme for Partitioning
  Irregular Graphs}.
\newblock \emph{SIAM Journal on Scientific Computing}, 20\penalty0
  (1):\penalty0 359--392, 1999.

\bibitem[Khalil et~al.(2008{\natexlab{a}})Khalil, Assadihaghi, Bila,
  Baillargeat, Aubourg, Verdeyme, Puech, and Lapierre]{Khalil2008a}
H.~Khalil, A.~Assadihaghi, S.~Bila, D.~Baillargeat, M.~Aubourg, S.~Verdeyme,
  J.~Puech, and L.~Lapierre.
\newblock {TOPOLOGY GRADIENT OPTIMIZATION IN 2-D AND 3-D FOR THE DESIGN OF
  MICROWAVE COMPONENTS}.
\newblock \emph{MICROWAVE AND OPTICAL TECHNOLOGY LETTERS}, 50\penalty0
  (10):\penalty0 895--896, 2008{\natexlab{a}}.
\newblock \doi{10.1002/mop}.

\bibitem[Khalil et~al.(2008{\natexlab{b}})Khalil, Delhote, Bila, Aubourg,
  Verdeyme, Puech, Lapierre, Delage, and Chartier]{Khalil2008b}
H.~Khalil, N.~Delhote, S.~Bila, M.~Aubourg, S.~Verdeyme, J.~Puech, L.~Lapierre,
  C.~Delage, and T.~Chartier.
\newblock {Topology optimization applied to the design of a dual-mode filter
  including a dielectric resonator}.
\newblock In \emph{Microwave Symposium Digest, 2008 IEEE MTT-S International},
  pages 1381--1384, 2008{\natexlab{b}}.
\newblock \doi{10.1109/MWSYM.2008.4633035}.

\bibitem[Khalil et~al.(2009)Khalil, Bila, Aubourg, Baillargeat, Verdeyme,
  Puech, Lapierre, Delage, and Chartier]{Khalil2009}
H.~Khalil, S.~Bila, M.~Aubourg, D.~Baillargeat, S.~Verdeyme, J.~Puech,
  L.~Lapierre, C.~Delage, and T.~Chartier.
\newblock {Topology optimization of microwave filters including dielectric
  resonators}.
\newblock In \emph{Microwave Conference, 2009. EuMC 2009. European}, pages
  687--690, sep 2009.

\bibitem[{Lazarov, Boyan Stefanov and Sigmund}(2010)]{Lazarov2010f}
O.~{Lazarov, Boyan Stefanov and Sigmund}.
\newblock {Filters in topology optimization based on Helmholtz-type
  differential equations}.
\newblock \emph{International Journal for Numerical Methods in Engineering},
  pages 1885--1891, 2010.
\newblock ISSN 0743-1619.
\newblock \doi{10.1002/nme}.

\bibitem[Nomura et~al.(2013)Nomura, Ohkado, Schmalenberg, Lee, Ahmed, and
  Bakr]{Nomura2013}
T.~Nomura, M.~Ohkado, P.~Schmalenberg, J.~Lee, O.~Ahmed, and M.~Bakr.
\newblock {Topology Optimization Method for Microstrips using Boundary
  Condition Representation}.
\newblock \emph{Proceedings of the 43rd European Microwave Conference}, pages
  632--635, 2013.

\bibitem[Ouedraogo et~al.(2013)Ouedraogo, Rothwell, Diaz, Fuchi, and
  Tang]{Quedraogo2012}
R.~O. Ouedraogo, E.~J. Rothwell, A.~R. Diaz, K.~Fuchi, and J.~Tang.
\newblock {Waveguide band-stop filter design using optimized pixelated
  inserts}.
\newblock \emph{MICROWAVE AND OPTICAL TECHNOLOGY LETTERS}, 55\penalty0 (1),
  2013.
\newblock \doi{10.1002/mop}.
\newblock URL \url{http://arxiv.org/abs/physics/0604155}.

\bibitem[Pozar(2005)]{Pozar2005}
D.~M. Pozar.
\newblock \emph{{Microwave Engineering}}.
\newblock John Wiley {\&} Sons, Ltd., 3rd edition, 2005.
\newblock URL
  \url{http://www.amazon.com/Microwave-Engineering-David-M-Pozar/dp/0470631554}.

\bibitem[Sigmund and Jensen(2003)]{Sigmund2003a}
O.~Sigmund and J.~S. Jensen.
\newblock {Systematic design of phononic band-gap materials and structures by
  topology optimization.}
\newblock \emph{Philosophical transactions. Series A, Mathematical, physical,
  and engineering sciences}, 361:\penalty0 1001--1019, 2003.
\newblock ISSN 1364-503X.
\newblock \doi{10.1098/rsta.2003.1177}.

\bibitem[Stolpe and Svanberg(2001)]{Stolpe2001}
M.~Stolpe and K.~Svanberg.
\newblock {An alternative interpolation scheme for minimum compliance topology
  optimization}.
\newblock \emph{Structural and Multidisciplinary Optimization}, 22\penalty0
  (2):\penalty0 116--124, 2001.
\newblock ISSN 1615147X.
\newblock \doi{10.1007/s001580100129}.

\bibitem[Svanberg(1987)]{Svanberg1987}
K.~Svanberg.
\newblock {The method of moving asymptotes - a new method for structural
  optimization}.
\newblock \emph{International Journal for Numerical Methods in Engineering},
  24\penalty0 (June 1986):\penalty0 359--373, 1987.
\newblock ISSN 1097-0207.
\newblock \doi{10.1002/nme.1620240207}.
\newblock URL \url{http://dx.doi.org/10.1002/nme.1620240207}.

\bibitem[Wang et~al.(2011)Wang, Lazarov, and Sigmund]{Wang2011a}
F.~Wang, B.~S. Lazarov, and O.~Sigmund.
\newblock {On projection methods, convergence and robust formulations in
  topology optimization}.
\newblock \emph{Structural and Multidisciplinary Optimization}, 43:\penalty0
  767--784, 2011.
\newblock ISSN 1615147X.
\newblock \doi{10.1007/s00158-010-0602-y}.

\end{thebibliography}







\end{document}